\documentclass[english]{article}
\usepackage[T1]{fontenc}
\usepackage{geometry}
\usepackage{amsmath}
\usepackage{setspace}
\usepackage{amssymb}

\makeatletter
\newcommand{\lyxaddress}[1]{
\par {\raggedright #1
\vspace{1.4em}
\noindent\par}
}

\usepackage{babel}
\makeatother

\begin{document}

\title{Super Symmetric Partners in $T^{4}$- space}

\author{P. S. Bisht and O. P. S. Negi}

\maketitle

\lyxaddress{\begin{center}
Department of Physics\\
Kumaun University\\
S. S. J. Campus\\
 Almora - 263601 (India)
\par\end{center}}

\begin{singlespace}
\begin{center}
Email: - ps\_bisht123@rediffmail.com
\par\end{center}

\begin{center}
ops\_negi@yahoo.co.in
\par\end{center}
\end{singlespace}

\begin{abstract}
Constructing the operators connecting the state of energy associated
with super partner Hamiltonians and super partner potentials for a
linear harmonic oscillator has been discussed and it is shown that
any super symmetric eigen state of one of the super partner potentials
in $T^{4}$- space is paired in energy with a symmetric eigen state
of the other partner potential.
\end{abstract}
\begin{quotation}
KEY WORDS: - SUPERSYMMETRY, SUPERLUMINAL TRANSFORMATION

PACS NO: 14.80LY
\end{quotation}

\section{INTRODUCTION:}

~~~~~~~During the past few years, there has been continuing
interest \cite{key-1,key-2} in higher dimensional kinematical models
for proper and unified theory of subluminal (bradyon) and superluminal
(tachyon) objects \cite{key-3}. The problem of representation and
localization of extended particles and superluminal objects may be
solved only by the use of higher dimensional space. Several attempts
of extending special theory of relativity to superluminal realm in
the usual four - space \cite{key-4} led to controversies \cite{key-5,key-6,key-7}
and satisfactory theory for tachyons could not be made acceptable
so far. Several experimental investigations \cite{key-8,key-9,key-10,key-11,key-12}
have been shown towards the evidences for the existence of tachyons:
particle moving faster than light $(v>c)$ (superluminal particles).
Still there are doubts (due to lack of experimental verification)\cite{key-13}
for the existence of tachyons and it still needs the necessity of
constructing a self-consistent quantum field theory, which might yield
their quantum properties relevant for their production and detection.
The continuing interest of tachyons showed that these particles are
not in contradiction to the special theory of relativity. A basic
disagreement in the various models of tachyons was the spin - statistics
relationship. Tanaka \cite{key-14}, Dhar - Sudarshan \cite{key-15},
Aron - Sudarshan \cite{key-16} assumed the spin for tachyons as that
of bradyons. On the other hand, Feinberg \cite{key-17}, Hamamoto
\cite{key-18} showed that spin-statistics for tachyons is reversed
to that for bradyons. Rajput and coworkers \cite{key-19,key-20} described
a Lorentz invariant quantum field theory of tachyons of various spins
and showed that tachyons are not localized in space. The Scalar and
Spinor field theory of free tachyons \cite{key-21} shows that tachyons
are localized in time. The localization space for the description
of tachyons is $T^{4}$ - space where the role of space and time (and
that of momentum and energy) are interchanged on passing from bradyons
to tachyons. The initial values of tachyons lie on the hyper plane
$x=0$ while those for tachyons lie on $t=0$. The space $R^{4}$
- (i.e. the localization space for bradyons) and $T^{4}$ - (the time
representation space for tachyons) demonstrate the structural symmetry
between these two and directly lead to space-time duality between
superluminal and subluminal objects. Supersymmetry, i.e. the Fermi
- Bose symmetry \cite{key-22,key-23,key-24,key-25,key-26,key-27,key-28,key-29,key-30,key-31,key-32,key-33,key-34},
is one of the most fascinating discoveries in the history of Physics.
The local extension of supersymmetric theories \cite{key-35,key-36,key-37,key-38,key-39}
(i.e. super gravity) provides a natural framework for the unification
of fundamental interactions of elementary particles. It was believed
earlier \cite{key-17} that spin$-0$ tachyons are quantized only
with anti commuting relations but their localization in space creates
problem with Lorentz invariance. Now it is made clear that tachyons
are localized in time and their localization space is $T^{4}$ - space
which behaves as that of bradyons do in $R^{4}$ - space and as such
it is not possible to accelerate directly a particle from $R^{4}$
- space to $T^{4}$- space (or subluminal to superluminal) for which
we need superluminal Lorentz transformations (SLT's).

Keeping these facts in mind, in this paper we have undertaken the
study of the super partner Hamiltonians in $T^{4}$ - space, having
the identical energy spectrum, expect for the ground states, and all
the potentials exhibiting the shape invariance contain the exact solutions.
Constructing the operations connecting the states of same energy for
super partner Hamiltonians $H_{+}(t)$ and $H_{-}(t)$ and supersymmetric
partner potentials $V_{+}(t)$ and $V_{-}(t)$ in $T^{4}$ - space,
all bound state wave functions have been calculated from ground state
and shape invariant potentials are obtained for supersymmetric linear
harmonic oscillator. It has been shown that $n^{th}$ excited state
energy of $H_{+}^{T}$ is identical to $(n+1)^{th}$ excited state
energy of $H_{-}^{T}$ while any supersymmetric eigen state of one
of the super partner potentials in $T^{4}$ - space is paired in energy
with a symmetric eigen state of the other partner potential. We have
also constructed a Semi - Unitary Transformation (SUT) to obtain the
supersymmetric partner Hamiltonians for a one dimensional harmonic
oscillator and it has been demonstrated that under this transformation
in $T^{4}$ - space the supersymmetric partner $H_{+}^{T}$ loses
its ground state but its eigen functions constitute a complete orthonormal
set in a subspace of full Hilbert space.

\section{BOSONIC AND FERMIONIC PARTNERS IN $T^{4}$ - SPACE}

~~~~~~In order to over come the various problems associated
with superluminal Lorentz transformations (SLTs), six - dimensional
formalism \cite{key-40,key-41,key-42,key-43,key-44} of space time
is adopted with the symmetric structure of space and time having three
space and three time components of a six dimensional space-time. In
this formalism, a subluminal observer $\mathcal{O}$ in the usual
$R^{4}\equiv(\overrightarrow{r},\, t)$ space is surrounded by a neighbourhood
in which one measures the scalar time $|t|\equiv(t_{x}^{2}+t_{y}^{2}+t_{z}^{2})^{\frac{1}{2}}$
and spatial vector $\overrightarrow{r}=(x,y,z)$ out of six independent
coordinates $(x,\, y,\, z,\, t_{x},\, t_{y},\, t_{z})$ of the six
- dimensional space $R^{6}$. On passing from $R^{6}=(\overrightarrow{r},\overrightarrow{\, t})$
to $(R^{6})^{t}=(\overrightarrow{r},\overrightarrow{\, t})^{T}$(
where $T$ stands the tachyon) via imaginary SLT's, the usual of observer
in $R^{6}$ will appear as $T^{4}\equiv(t_{x},\, t_{y},t_{z},r)^{T}$
to the observer $\mathcal{O}$ in $R^{6}$. The resulting space for
bradyons and tachyons is thus identified as the $R^{6}$- or $M(3,\,3)$
space where both space and time as well as energy and momentum are
considered as vector quantities. Superluminal Lorentz transformations
(SLTs) between two frames $K$ and $K'$ moving with velocity $v>1$
are defined in $R^{6}$- or $M(3,\,3)$ space as follows;

\begin{eqnarray}
x' & \longrightarrow & \pm t_{x},\nonumber \\
y' & \longrightarrow & \pm t_{y},\nonumber \\
z' & \longrightarrow & \pm\gamma\,(z-vt),\nonumber \\
t_{x}' & \longrightarrow & \pm x,\nonumber \\
t_{y}' & \longrightarrow & \pm y,\nonumber \\
t_{z}' & \longrightarrow & \pm\gamma\,(t-vz),\label{eq:1}\end{eqnarray}
where $\gamma\,=\,(v^{2}-1)^{-\frac{1}{2}}.$ These transformations
lead to the mixing of space and time coordinates for transcendental
tachyonic objects, $(|\overrightarrow{v}|\rightarrow\infty)$ where
equation (\ref{eq:1}) takes the following form;

\begin{eqnarray}
+\,\, dt_{x}\rightarrow dt_{x}' & = & dx\,\,\,\,+,\nonumber \\
+\,\, dt_{y}\rightarrow dt_{y}' & = & dy\,\,\,\,+,\nonumber \\
+\,\, dt_{z}\rightarrow dt_{z}' & = & dz\,\,\,\,+,\nonumber \\
-\,\, dz\rightarrow dz' & = & dt_{z}\,\,\,\,-,\nonumber \\
-\,\, dy\rightarrow dy' & = & dt_{y}\,\,\,\,-,\nonumber \\
-\,\, dx\rightarrow dx' & = & dt_{x}\,\,\,\,-.\label{eq:2}\end{eqnarray}
It shows that we may have only two four dimensional slices of $R^{6}$-
or $M(3,\,3)$ space $(+,+.+,-)$ and $(-,-,-,+)$. When any reference
frame describes bradyonic objects it is necessary to describe $M(1,3)\Longrightarrow[t,\, x,\, y,\, z]\,(R^{4}-space)$
so that the coordinates $t_{x}$ and $t_{y}$ are not observed or
couple together giving $t=(t_{x}^{2}+t_{y}^{2}+t_{z}^{2})^{½}$.
On the other hand when a frame describes bradyonic object in frame
$K$, it will describe a tachyonic object (with velocity $|\overrightarrow{v}|\rightarrow\infty$
) in $K'$ with $M'(1,\,3)$ space i.e.$\,\, M'(1,\,3)=[t_{z}',\, x',\, y',\, z']=[z,\, t_{x},\, t_{y},\, t_{z}](T^{4}-space).$

As such, we define $M'(1,\,3)$ space as $T^{4}$- space or $M(3,\,1)$
space where $x$ and $y$ are not observed but coupled together giving
rise to $r=(x^{2}+y^{2}+z^{2})^{1/2}$. As such, the spaces $R^{4}$
and $T^{4}$ are two observational slices of $R^{6}$- or $M(3,\,3)$
space. Unfortunately the space $R^{6}-or\, M(3,\,3)$ is not consistent
with special theory of relativity and accordingly the subluminal and
superluminal Lorentz transformations lose their meaning in $R^{6}$-
or $M(3,3)$ space in the sense that these transformations do not
represent either the bradyonic or tachyonic objects in this space.
It has been shown earlier \cite{key-2,key-3} that the true localizations
space for bradyons is $R^{4}$ - space while that for tachyons is
$T^{4}$ - space. So a bradyonic $R^{4}=M(1,3)$ space now maps to
a tachyonic $T^{4}=M'(3,1)$ space or vice versa i.e. 

\begin{eqnarray}
R^{4}=M(1,3) & \overset{SLT}{\rightarrow} & M'(3,1)=T^{4}.\label{eq:3}\end{eqnarray}
Hence $R^{4}$, the suitable space for describing bradyonic phenomena,
is mapped under SLT's onto the space $T^{4}$ which is the space suitable
to describe the tachyonic phenomena. In such formalism we get the
following mapping for the position four - vector $\{x_{\mu}\}=(\overrightarrow{r},-t)$,
four - differential $\{\nabla_{\mu}\}=(\overrightarrow{\nabla},-\partial_{t})$
and four - potential $\{A_{\mu}\}=(\overrightarrow{A},-\phi_{e})$
under SLT's as 

\begin{eqnarray*}
\left\{ \overrightarrow{r},-t=(t_{x}^{2}+t_{y}^{2}+t_{z}^{2})^{\frac{1}{2}}\right\}  & \longrightarrow & \left\{ \overrightarrow{t},-r=(x^{2}+y^{2}+z^{2})^{\frac{1}{2}}\right\} ^{T}\end{eqnarray*}

\begin{eqnarray}
\left\{ \overrightarrow{\nabla_{r}},-\partial_{t}=\partial_{t_{x}}^{2}+\partial_{t_{y}}^{2}+\partial_{t_{z}}^{2}\right\}  & \longrightarrow & \left\{ \overrightarrow{\nabla_{t}},-\partial_{r}=\partial_{x}^{2}+\partial_{y}^{2}+\partial_{z}^{2}\right\} ^{T}\nonumber \\
\left\{ \overrightarrow{\{A},-\phi_{e}=\phi_{ex}^{2}+\phi_{ey}^{2}+\phi_{ez}^{2}\right\}  & \longrightarrow & \left\{ \overrightarrow{\phi_{e}},-A=(A_{x}^{2}+A_{y}^{2}+A_{z}^{2})^{\frac{1}{2}}\right\} ^{T}\label{eq:4}\end{eqnarray}
where $\overrightarrow{\nabla}_{r}$ are $\overrightarrow{\nabla}_{t}$
differential operators in terms of three spatial and three temporal
coordinates, respectively. Applying these mappings on the equations
for generalized electric field, magnetic field and Lorentz force associated
with dyons , we get the following expressions respectively for generalized
electric field, magnetic field and Lorentz force on dyons under superluminal
transformations in $T^{4}$ - space i.e.;

\begin{eqnarray}
\overrightarrow{E} & ^{T}=- & \frac{\partial\overrightarrow{\phi}_{e}}{\partial r}-\overrightarrow{\nabla}_{t}A+(\overrightarrow{\nabla}_{t}\times\overrightarrow{\phi}_{g})\nonumber \\
\overrightarrow{H}\,^{T} & = & \frac{\partial\overrightarrow{\phi}_{g}}{\partial r^{t}}+\overrightarrow{\nabla}_{t}B+(\overrightarrow{\nabla}_{t}\times\overrightarrow{\phi}_{e})\nonumber \\
\overrightarrow{F}\,^{T} & = & Real\, q*[\overrightarrow{\psi}*-(\overleftarrow{u}\times\overrightarrow{\psi}*)\label{eq:5}\end{eqnarray}
where $\{B_{\mu}\}=\{\overrightarrow{B},-\phi_{g}\}$ is introdued
\cite{key-45} as magnetic four - potential beside the electric four
- potential $\{A_{\mu}\};$ and $\overrightarrow{\psi}=\overrightarrow{E}-i\overrightarrow{H}$
is described as generalized electromagnetic vector field in subluminal
frame of reference ; and $\overleftarrow{u}=\frac{d\overleftarrow{t}'}{dr'}$
is inverse velocity. These equations show the superluminal fields
associated with tachyonic dyons containing both longitudinal and transverse
parts. Equation $\overrightarrow{F}\,^{T}=Real\, q*[\overrightarrow{\psi}*-(\overleftarrow{u}\times\overrightarrow{\psi}*)$
describes the Lorentz force acting on dyon in $T^{4}$-space. It may
be recalled as the force of tachyonic dyon which is similar to that
of dyonic force in $R^{4}$- space if we replace the particle velocity
$\overrightarrow{u}=\frac{d\overrightarrow{r}}{dt}$ to that of the
inverse velocity $\overleftarrow{u}=\frac{d\overleftarrow{t}'}{dr'}$
alon with the role of space variables is interchanged with time variables.
So, the theories of tachyons be better understood in $T^{4}$- space
where they behave as bradyons do in $R^{4}$- space. As such in view
of their localizability and other quantum properties, we can formulate
the supersymmetric theories of tachyons in $T^{4}$- space in the
parallel ground of bradyonic supersymmetric theories in $R^{4}$-
space . Similarly, the total Hamiltonian $H^{T}$ of a tachyonic supersymmetric
system may be decomposed in to its bosonic part $H_{B}^{T}$ and the
fermionic part $H_{F}^{T}$ such that

\begin{eqnarray}
H^{T} & = & H_{B}^{T}+H_{F}^{T}\label{eq:6}\end{eqnarray}
with

\begin{eqnarray}
H_{B}^{T} & = & -\frac{E^{2}}{2k}+\frac{1}{2}\{W'(t)\}^{2}\label{eq:7}\end{eqnarray}
and

\begin{eqnarray}
H_{F}^{T} & = & \frac{1}{2}W''(t)[\overline{\psi_{T}},\psi_{T}]=-iW''(t)Y_{T}\label{eq:8}\end{eqnarray}
where$k$is the rest mass of tachyon, $W(t)$ is super potential while
$\psi_{T}$ and $\overline{\psi_{T}}$ are the fermionic variables
in $T^{4}$- space describing spin degree of freedom and the operator
$Y_{T}$ is defined as 

\begin{eqnarray}
Y_{T} & = & \frac{i}{2}\{\overline{\psi}_{T},\psi_{T}\}.\label{eq:9}\end{eqnarray}
In ( \ref{eq:8}), prime and double prime denote first and second
order derivatives. Supersymmetric Hamiltonian may also be constructed
in terms of non - Hermitian supercharge operators $Q$ and $Q^{+}$
as 

\begin{eqnarray}
H^{T} & =\frac{1}{2}\left\{ \, Q,\, Q^{+}\right\}  & =\frac{1}{2}\,\left[QQ^{+}+Q^{+}Q\right]\label{eq:10}\end{eqnarray}
such that

\begin{eqnarray}
\left[Q,Q^{+}\right] & = & 0,\,\,\,\, Q^{2}=Q^{+^{2}}=0\label{eq:11}\end{eqnarray}
and

\begin{eqnarray}
\left[H^{T},\, Q\right] & =\left[H^{T},Q^{+}\right]= & 0;\nonumber \\
\left[Y_{T}\,,Q\right] & = & -iQ;\nonumber \\
\left[Y_{T},\, Q^{+}\right] & = & iQ^{+}.\label{eq:12}\end{eqnarray}
From equations (\ref{eq:6}) and (\ref{eq:10}), the supercharges
may readily be constructed accordingly in the following form;

\begin{eqnarray}
Q^{+} & = & \left[E-iW'(t)\right]\overline{\psi_{T}};\,\,\,\, Q=\left[E+iW'(t)\right]\psi_{T}.\label{eq:13}\end{eqnarray}
Any state $\left|B^{T}\right\rangle $ satisfies the conditions

\begin{eqnarray}
Q\left|B^{T}\right\rangle  & = & 0;\,\,\,\, Q^{+}\left|B^{T}\right\rangle \neq0\label{eq:14}\end{eqnarray}
is denoted as the bosonic state for which we have $H^{T}\left|B^{T}\right\rangle =\frac{1}{2}QQ^{+}\left|B^{T}\right\rangle .$
Similarly, fermionic state $\left|F^{T}\right\rangle $ satisfies
the conditions

\begin{eqnarray}
Q^{+}\left|F^{T}\right\rangle  & = & 0;\,\,\,\,\,\, Q\left|F^{T}\right\rangle \neq0\label{eq:15}\end{eqnarray}
which gives $H^{T}\left|F^{T}\right\rangle =\frac{1}{2}Q^{+}Q\left|F^{T}\right\rangle .$
Using these relations, it may readily be demonstrated that the operator
$Q$ transforms states $\left|F^{T}\right\rangle $ in to states $\left|B^{T}\right\rangle $
of the same eigen energy $E^{T}$ and the operator $Q^{+}$ transforms
the states $\left|B^{T}\right\rangle $ into states $\left|F^{T}\right\rangle $,
i.e.

\begin{eqnarray}
Q\,\left|F^{T}\right\rangle  & = & E^{T^{\frac{1}{2}}}\left|F^{T}\right\rangle ;\,\,\,\, Q^{+}\left|F^{T}\right\rangle =E^{T^{\frac{1}{2}}}\left|F^{T}\right\rangle \label{eq:16}\end{eqnarray}
which directly shows $\left[Q,\, Q^{+}\right]=0,\,\,\,\, Q^{2}=Q^{+^{2}}=0$;
the supersymmetry between tachyonic fermions and tachyonic bosons
with positive momentum eigen values while the negative momentum value
corresponds to singularity problem associated with super potential
in this supersymmetric model.

Alternatively, supersymmetric quantum mechanics may be worked out
in terms of a pair of bosonic Hamiltonians $H_{-}^{T}$ and $H_{+}^{T}$
which are supersymmetric partners of supersymmetric Hamiltonian i.e.,

\begin{eqnarray}
H^{T} & = & H_{-}^{T}\bigoplus H_{-}^{T}.\label{eq:17}\end{eqnarray}
In order to construct these super partner Hamiltonians for the system
described by equation (\ref{eq:6}), let us introduce the potential
$V_{-}(t)$ whose ground state energy has been adjusted to zero with
the corresponding ground state wave function given by;

\begin{eqnarray}
\psi_{0}^{T(-)} & = & \exp[-\int_{0}^{t}W(t')dt'].\label{eq:18}\end{eqnarray}
Substituting it in the Schrödinger equation (in the units of $\hbar=2k=1$
), we get

\begin{eqnarray}
V_{-}(t) & = & W^{2}(t)-W'(t)=\frac{\psi_{0}^{T''(-)}}{\psi_{0}^{(-)}}.\label{eq:19}\end{eqnarray}
Hence we have the following Hamiltonian corresponding to the potential,

\begin{eqnarray}
H_{-}^{T} & =-\frac{d^{2}}{dt^{2}}+\frac{\psi_{0}^{T''(-)}}{\psi_{0}^{(-)}} & =-\frac{d^{2}}{dt^{2}}+V_{-}(t).\label{eq:20}\end{eqnarray}
If the ground state wave function $\psi_{0}^{T(-)}$ is square integrable
then the supersymmetry will be broken. This Hamiltonian may also be
written in the following form in terms of bosonic operator $\hat{B}$
and $\hat{B^{+}}$;

\begin{eqnarray}
H_{-}^{T} & = & \hat{B^{+}}\hat{B}\label{eq:21}\end{eqnarray}
where

\begin{eqnarray}
\hat{B} & =\frac{d}{dt}+W(t) & =\frac{d}{dt}-\frac{\psi_{0}^{T''(-)}}{\psi_{0}^{(-)}};\nonumber \\
\hat{B^{+}} & =-\frac{d}{dt}+W(t) & =-\frac{d}{dt}-\frac{\psi_{0}^{T''(-)}}{\psi_{0}^{(-)}}.\label{eq:22}\end{eqnarray}
Let us introduce the Hamiltonian 

\begin{eqnarray}
H_{+}^{T} & =\hat{B^{+}}\hat{B}=-\frac{d^{2}}{dt^{2}}+V_{+}(t) & =-\frac{d^{2}}{dt^{2}}+W^{2}(t)-W'(t)\label{eq:23}\end{eqnarray}
where

\begin{eqnarray}
V_{+}(t) & =W^{2}(t)-W'(t) & =V_{-}(t)+2W'(t).\label{eq:24}\end{eqnarray}
The potentials $V_{+}$ and $V_{-}$ are called supersymmetric partner
potentials and $H_{+}$ is the Hamiltonian corresponding to the potential
$V_{+}(t)$. From equations (\ref{eq:13}) and (\ref{eq:18}), we
get

\begin{eqnarray}
\frac{1}{2}\left[V_{+}(t)+V_{-}(t)\right] & = & W^{2}(t);\nonumber \\
\left[\hat{B},\hat{B^{+}}\right] & = & 2W'(t)\label{eq:25}\end{eqnarray}
showing that $W^{2}(t)$ is the average of the potential $V_{+}(t)$
and $V_{-}(t)$, whereas $W'(t)$ is proportional to the commutator
of $\hat{B}$ and $\hat{B^{+}}$. The supersymmetric charges have
been defined in terms of operators $\hat{B^{+}}$and $\hat{B}$. Hence,
the Hamiltonians $H_{+}^{T}$ and $H_{-}^{T}$ are denoted as supersymmetric
partners. It can be visualized by introducing supersymmetric charges
$Q$ and $Q^{+}$ as

\begin{eqnarray}
Q & = & \sqrt{2}\left[\begin{array}{cc}
0 & 0\\
B & 0\end{array}\right];\,\,\, Q^{+}=\sqrt{2}\left[\begin{array}{cc}
0 & B^{+}\\
0 & 0\end{array}\right].\label{eq:26}\end{eqnarray}
Then we have

\begin{eqnarray*}
H_{SUSY}^{T} & =\frac{1}{2}\left\{ \, Q,\, Q^{+}\right\}  & =\left[\begin{array}{cc}
H_{-} & 0\\
0 & H_{+}\end{array}\right]\end{eqnarray*}
or

\begin{eqnarray*}
\left[H_{SUSY}^{T},Q\right] & =\left[H_{SUSY}^{T},Q^{+}\right] & =0\end{eqnarray*}
or

\begin{eqnarray}
Q^{2}=Q^{+^{2}} & = & 0.\label{eq:27}\end{eqnarray}
For any eigen function $\psi^{T(-)}$ of $H_{-}^{T}$ with the corresponding
eigen value $E^{T}$, we have

\begin{eqnarray*}
\hat{B^{+}}\hat{B}\psi^{T(-)} & = & E^{T}\psi^{T(-)}\end{eqnarray*}
or

\begin{eqnarray}
H_{+}^{T}[\hat{B}\psi^{T(-)}] & = & E^{T}\hat{[B}\psi^{T(-)}]\label{eq:28}\end{eqnarray}
which shows that $\hat{B}\psi^{T(-)}$ is the eigen function of $H_{+}^{T}$,
with the same eigen value $E^{T}$. Similarly it can be demonstrated
that if $\psi^{T(+)}$ is an eigen function of $H_{+}^{T}$ with the
eigen value $E^{'T}$ then $\hat{B^{+}}\psi^{T(-)}$ is an eigen function
of $H_{-}^{T}$ with the same eigen value. Thus $H_{-}^{T}$ and $H_{+}^{T}$
have the identical energy spectrum except for the ground state of
$H_{-}^{T}$. It is also obvious that the operators $\hat{B}$ and
$\hat{B^{+}}$ connect the states of the same energy for two different
supersymmetric partner potentials $V_{+}(t)$ and $V_{-}(t)$. Specific
relations between spectra of $H_{-}$ and $H_{+}$ are based on the
symmetry between the solvable potential and its supersymmetric partner
potential.

\section{SUPERSYMMETRY PARTNERS IN $T^{4}$ - SPACE}

For supersymmetry one - dimensional Harmonic oscillator we have the
super potential

\begin{eqnarray}
W(t) & = & \Omega t.\label{eq:29}\end{eqnarray}
Then the supersymmetic hamiltonian is described as 

\begin{eqnarray}
H_{-}^{T} & = & \frac{d^{2}}{dt^{2}}+\Omega^{2}t^{2}-\Omega\label{eq:30}\end{eqnarray}
and equations (\ref{eq:19}) and (\ref{eq:20}) give

\begin{eqnarray}
V_{-}(t) & = & \Omega^{2}t^{2}-\Omega.\label{eq:31}\end{eqnarray}
Similarly, equation (\ref{eq:28}) leads to

\begin{eqnarray}
V_{+}(t) & = & \Omega^{2}t^{2}+\Omega.\label{eq:32}\end{eqnarray}
From equation (\ref{eq:22}), we may then readily get

\begin{eqnarray}
B^{+} & = & -\frac{d}{dt}+\Omega t;\nonumber \\
B^{-} & = & \frac{d}{dt}+\Omega t\,\,\,\,\,\label{eq:33}\end{eqnarray}
which gives

\begin{eqnarray}
H_{+}^{T} & =- & \frac{d^{2}}{dt^{2}}+\Omega^{2}t^{2}+\Omega.\label{eq:34}\end{eqnarray}
Substituting relation (\ref{eq:22}) into equation (\ref{eq:18}),
we get

\begin{eqnarray}
\psi_{0}^{T(-)} & = & \left(\frac{\Omega}{\pi}\right)^{\frac{1}{4}}\exp[-\frac{\Omega t^{2}}{2}]\label{eq:35}\end{eqnarray}
where the constant $\left(\frac{\Omega}{\pi}\right)^{\frac{1}{4}}$
is the result of the orthonormality of ground state wave function.
From equation (\ref{eq:30}) and (\ref{eq:34}), it is obvious that 

\begin{eqnarray*}
H_{-}^{T}\psi_{0}^{T(-)} & = & 0\end{eqnarray*}
i.e.

\begin{eqnarray}
H_{0}^{T(-)} & = & 0.\label{eq:36}\end{eqnarray}
Using relations (\ref{eq:34}) and (\ref{eq:35}), we get

\begin{eqnarray}
H_{+}^{T}\psi_{0}^{T(-)} & =2\Omega\psi_{0}^{T(-)}(t) & =\Omega'\psi_{0}^{T(-)}(t)\label{eq:37}\end{eqnarray}
where $\Omega'=2\Omega$ is the classical frequency of the oscillator.
This equation shows that ground state of $H_{+}^{T}$ is not the zero
energy state i.e. $E_{0}^{T(+)}=\Omega'$.

First excited state $\psi_{1}^{T(-)}$ of $H_{-}^{T}$ may be obtained
by the raising operator $B^{+}$ i.e.

\begin{eqnarray}
\psi_{1}^{T(-)}(t) & = & B^{+}\psi_{0}^{T(-)}(t)\nonumber \\
= & \left[-\frac{d}{dt}+\Omega t\right] & \left(\frac{\Omega}{\pi}\right)^{\frac{1}{4}}\exp\left[-\frac{\Omega t^{2}}{2}\right]\nonumber \\
= & 2\left(\frac{\Omega}{\pi}\right)^{\frac{1}{4}}\Omega t\, & \exp\left[-\frac{\Omega t^{2}}{2}\right].\label{eq:38}\end{eqnarray}
Then we get

\begin{eqnarray}
H_{-}^{T}\psi_{1}^{T(-)} & = & \Omega'\psi_{1}^{T(-)}(t)\label{eq:39}\end{eqnarray}
i.e.

\begin{eqnarray*}
E_{1}^{T(-)} & = & \Omega'.\end{eqnarray*}
Equations (\ref{eq:38}) and (\ref{eq:39}) yield 

\begin{eqnarray}
E_{0}^{T(+)} & = & E_{1}^{T(-)}\label{eq:40}\end{eqnarray}
i.e. ground state energy of $H_{+}^{T}$is the first excited state
energy of $H_{-}^{T}$. Similarly, we get

\begin{eqnarray}
H_{+}^{T}\psi_{1}^{T(-)}(t) & = & 2\Omega'\psi_{1}^{T(-)}(t)\label{eq:41}\end{eqnarray}
i.e.

\begin{eqnarray*}
E_{1}^{T(+)} & = & 2\Omega'.\end{eqnarray*}
Second excited state of $H_{-}^{T}$ may be constructed as follows;

\begin{eqnarray}
\psi_{2}^{T(-)}(t) & = & \frac{B^{+}}{2}\psi_{1}^{T(-)}(t)=2\left(\frac{\Omega}{\pi}\right)^{\frac{1}{4}}\Omega[-1+2\Omega t^{2}]\exp[-\frac{\Omega t^{2}}{2}]\label{eq:42}\end{eqnarray}
for which we have 

\begin{eqnarray}
H_{-}^{T}\psi_{2}^{T(-)}(t) & = & 2\Omega'\psi_{2}^{T(-)}(t)\nonumber \\
E_{2}^{T(-)} & = & 2\Omega'\nonumber \\
H_{-}^{T}\psi_{2}^{T(-)}(t) & = & 3\Omega'\psi_{2}^{T(-)}(t)\label{eq:43}\end{eqnarray}
showing that first excited state energy of $H_{+}^{T}$ is state $\psi_{1}^{T(-)}$
is identical to the second excited energy of $H_{-}^{T}$ in the state
$\psi_{2}^{T(-)}$. Generalizing these results, it may be inferred
that with respect to the states $\psi_{n}^{T(-)}$ , the $n^{th}$
excited state energy of $H_{+}^{T}$ is identical to the $(n+1)^{th}$
excited state energy of $H_{-}^{T}$. In other words, any state $\psi_{n}^{T(-)}$
of the super partner Hamiltonian $H_{-}^{T}$ corresponding to the
eigen value $E_{n}^{T}$ of harmonic oscillator is also the eigen
state of $H_{+}^{T}$ with the corresponding eigen value $E_{n+1}^{T}$
.

Let us construct the eigen state $\psi_{n}^{T(+)}(t)$ of $H_{+}^{T}$
from the states $\psi_{n}^{T(-)}(t)$ of $H_{-}^{T}$ in the following
form;

\begin{eqnarray}
\psi_{n}^{T(+)}(t) & = & \frac{1}{\sqrt{E_{n+1}^{T(-)}}}B\psi_{n+1}^{T(-)}(t);\nonumber \\
\psi_{0}^{T(+)}(t) & = & \frac{1}{\sqrt{E_{1}^{T(-)}}}\left(\frac{d}{dt}+\Omega t\right)\psi_{1}^{T(-)}(t)\label{eq:44}\end{eqnarray}
which may be written in the following form by using equation (\ref{eq:18});

\begin{eqnarray}
\psi_{0}^{T(+)}(t) & = & \left(\frac{\Omega}{\pi}\right)^{\frac{1}{4}}\sqrt{2\Omega}\exp[-\frac{\Omega t^{2}}{2}].\label{eq:45}\end{eqnarray}
Then we get,

\begin{eqnarray}
H_{+}^{T}\psi_{0}^{T(+)} & = & E_{0}^{T(+)}\Omega'\psi_{0}^{T(+)}(t);\,\,\,\,\,=\Omega'=E_{1}^{T(-)}.\label{eq:46}\end{eqnarray}
Similarly, we get

\begin{eqnarray}
\psi_{1}^{T(+)}(t) & = & \left(\frac{\Omega}{\pi}\right)^{\frac{1}{4}}3(\Omega)^{\frac{3}{2}}t\exp[-\frac{\Omega t^{2}}{2}];\nonumber \\
\psi_{2}^{T(+)}(t) & = & 2\left(\frac{\Omega}{\pi}\right)^{\frac{1}{4}}(\Omega)^{\frac{3}{2}}[-1+2\Omega t^{2}]\exp[-\frac{\Omega t^{2}}{2}];\label{eq:47}\end{eqnarray}
which gives rise to

\begin{eqnarray}
H_{+}^{T}\psi_{1}^{T(+)} & = & 2\Omega'\psi_{1}^{T(+)}(t);\,\,\,\,\,\,\, E_{1}^{T(+)}=2\Omega'=E_{2}^{T(-)}\label{eq:48}\end{eqnarray}
and

\begin{eqnarray}
H_{+}^{T}\psi_{2}^{T(+)} & =3 & \Omega'\psi_{2}^{T(+)}(t):\,\,\,\,\,\,\: E_{2}^{T(+)}=3\Omega'=E_{3}^{T(-)}.\label{eq:49}\end{eqnarray}
Generalizing these results, we may infer that\begin{eqnarray}
E_{n}^{T(+)} & = & E_{n+1}^{T(-)}\label{eq:50}\end{eqnarray}
which shows that $n^{th}$ excited state energy of $H_{+}^{T}$ is
identical to $(n+1)^{th}$ excited state energy of $H_{-}^{T}$ .
In other words, we may infer that if $\psi_{n}^{T(-)}$ is the $n^{th}$
eigen state of $H_{-}^{T}$ with the corresponding eigen value $E_{n}^{T(+)}$
then $\frac{1}{\sqrt{E_{n}^{T(-)}}}B\psi_{n}^{T(-)}(t)$ is the $(n-1)^{th}$
eigen state of $H_{+}^{T}$ with the energy $E_{n}^{T(+)}$. It is
obvious from equations (\ref{eq:34}) and (\ref{eq:45}); (\ref{eq:37}),
(\ref{eq:38}) and (\ref{eq:42}); and (\ref{eq:47}) that for supersymmetry
harmonic oscillator we have

\begin{eqnarray}
\psi_{n}^{T(+)} & \propto & \psi_{n}^{T(-)}\label{eq:51}\end{eqnarray}
with the corresponding energy eigen values satisfying condition (\ref{eq:50}).
From equations (\ref{eq:31}) and (\ref{eq:32}) we get the following
conditions for the supersymmetry partner potentials for one - dimensional
harmonic oscillator

\begin{eqnarray}
V_{\pm}(t) & = & V_{\pm}(-t).\label{eq:52}\end{eqnarray}
Any anti - symmetric eigen state of one of these potentials is paired
in energy with the symmetric eigen state of the other equations (\ref{eq:34}),
(\ref{eq:42}), (\ref{eq:45}) and (\ref{eq:47}) shows that all eigen
functions of $H_{+}^{T}$ and $H_{-}^{T}$ will go to non-zero constant
as $t\rightarrow0$ and hence these are unacceptable as physical states
of $V_{\pm}(t)$ . These unacceptable solutions of both these super
partner potentials are paired in energy and hence the degeneracy theorem
holds for $V_{\pm}(t)$ maintaining the supersymmetry of the system.

\section{DISCUSSION}

Equation (\ref{eq:5}) for Lorentz force acting on dyon interacting
with generalized superluminal electromagnetic fields in $T^{4}$ -
space is similar to that of a dyon interacting with generalized subluminal
electromagnetic fields in $R^{4}$- space except that the role of
velocity $\overrightarrow{u}=\frac{d\overrightarrow{r}'}{dt}$ has
been changed with the inverse velocity $\overleftarrow{u}=\frac{d\overrightarrow{t}'}{dr}$
and consequently the role of space and time variables are interchanged
on passing from subluminal to superluminal realm via SLT's and thus
suggests that tachyonic dyons are localized in time with their fields
having rotational symmetry in temporal planes and their motion is
bi - dimensional in time.

It is emphasized that there are two observations slices, $R^{4}$-
or $M(1,3)$ (one time and three space) and $T^{4}$- or $M(3,1)$
(three time and one space), in the existing $R^{6}$-space, $M(3,3)$
(three time and three space). The transformations for transcendent
velocity $(v\rightarrow\infty)$ reveal that the tachyons and the
bradyons cannot be described individually in the $R^{6}$ - space.
As such, the equivalence of space and time and the necessity of a
four-dimensional Minkowski space to describe any physical event consistent
with the special theory of relativity, the $T^{4}$ - space have been
constructed as the natural space for the description of the tachyons,
whereas a bradyon is described in the usual $R^{4}$- space. It has
been stressed that a bradyonic $R^{4}$- space maps under SLT's to
a tachyonic $T^{4}$ - space and on passing from a bradyon to a tachyon
via SLT's, the $T^{4}$ - space becomes the representation space of
the observables while the $R^{4}$- space represents the internal
space, where only internal degree for bradyons are represented. It
means that the usual pseudo - Euclidean group, $E_{3}$ of space -
time representation of the Poincare group will be visualized as the
usual rotation group in the three time planes in the $T^{4}$- space.
Further an object moving with velocity, $v<1$ (bradyonic) in subspace
$R^{4}$ (forward velocity $v=\frac{dr}{dt}$ ) appears to be moving
with an inverse velocity $\left(u=\frac{dt}{dr}\right)$ with $u>1$.
It may also be emphasized that for all-purpose the superluminal fields
(tachyonic) in the time-energy representation ($T^{4}$- space) behave
as subluminal (bradyonic) fields in space-momentum representation
($R^{4}$- space). Equation (\ref{eq:6} - \ref{eq:8}) shows the
total supersymmetric Hamiltonian in $T^{4}$- space associated with
tachyons in terms of energy eigen values and super potentials. Equation
(\ref{eq:16}) shows that the supersymmetry between tachyonic fermions
and tachyonic bosons with positive momentum eigen values while the
negative momentum values corresponds to singularity problem associated
with super potential in the supersymmetric model. Constructing the
potential $V_{-}(t)$ in the form given by equation (\ref{eq:19})
with the ground state energy vanishing and the ground state wave function
given by equation (\ref{eq:18}) the super - partner Hamiltonian $H_{-}^{T}$
has been obtained in the form of equation (\ref{eq:20}) and it has
been found that its ground state energy identical to ground state
momentum of a bradyon is vanishing provided that the wave - function
defined by equation (\ref{eq:19}) is square integrable.

The Hamiltonian $H_{+}^{T}$ (super - partner potential of $H_{-}^{T}$
) has been constructed in the form of equation (\ref{eq:23}) in terms
of potential $V_{+}(t)$ (super-partner potential of $V_{-}(t)$ )
given by equation (\ref{eq:24}). The supersymmetric charge operators
are introduced by equation (\ref{eq:26}) in $T^{4}$- space. It has
been shown that $H_{+}^{T}$ and $H_{-}^{T}$ have identical energy
(momentum) spectrum in $T^{4}(R^{4})$ - space except for the ground
state of a Hamiltonian. Equation (\ref{eq:41}) and (\ref{eq:43})
shows that the first excited state energy of $H_{+}^{T}$ in state
$\psi_{1}^{T(-)}$ is identical to the second excited energy of $H_{-}^{T}$
in the state $\psi_{2}^{T(-)}$. Generalization of these results leads
to the inference that with respect to the states $\psi_{n}^{T(-)}$
the $n^{th}$ excited state energy of $H_{+}^{T}$ is identical to
the $(n+1)^{th}$ excited state energy of $H_{-}^{T}$. In other words
any state $\psi_{n}^{T(-)}$ of the super partner Hamiltonian $H_{-}^{T}$
corresponding to the eigen value $E_{n}^{T}$ of harmonic oscillator
is also the eigen state of $H_{+}^{T}$with the corresponding eigen
value $E_{n+1}^{T}$. Equations (\ref{eq:34}), (\ref{eq:42}), (\ref{eq:45})
and (\ref{eq:47}) provide that all eigen functions of $H_{+}^{T}$
and $H_{-}^{T}$ will go to non - zero constant as $t\rightarrow0$
and hence these are unacceptable as physical states of $V_{\pm}(t)$
. These unacceptable solutions of both these super partner potentials
are paired in energy and hence the degeneracy theorem holds for $V_{\pm}(t)$
maintaining the supersymmetry of the system in $T^{4}$ - space.

\begin{quotation}
\textbf{Acknowledgement:} - This work is carried out under Uttarakhand
State Council for Science and Technology, Dehradun research scheme.
\end{quotation}

\end{document}